\RequirePackage[mathlines]{lineno} 
\documentclass[a4paper,11pt]{article}
\pdfoutput=1 

\usepackage{jheppub} 

\usepackage[T1]{fontenc} 

\usepackage{threeparttable}
\usepackage{multirow}
\usepackage{subfigure}
\usepackage{float}
\let\oldequation\equation
\let\oldendequation\endequation
\renewenvironment{equation}
  {\linenomathNonumbers\oldequation}
  {\oldendequation\endlinenomath}

\begin{document}
\setrunninglinenumbers

\title{\bf\boldmath Measurement of the branching fraction of $D^+ \to \tau^+\nu_{\tau}$}
\collaborationImg{\includegraphics[height=30mm, angle=90]{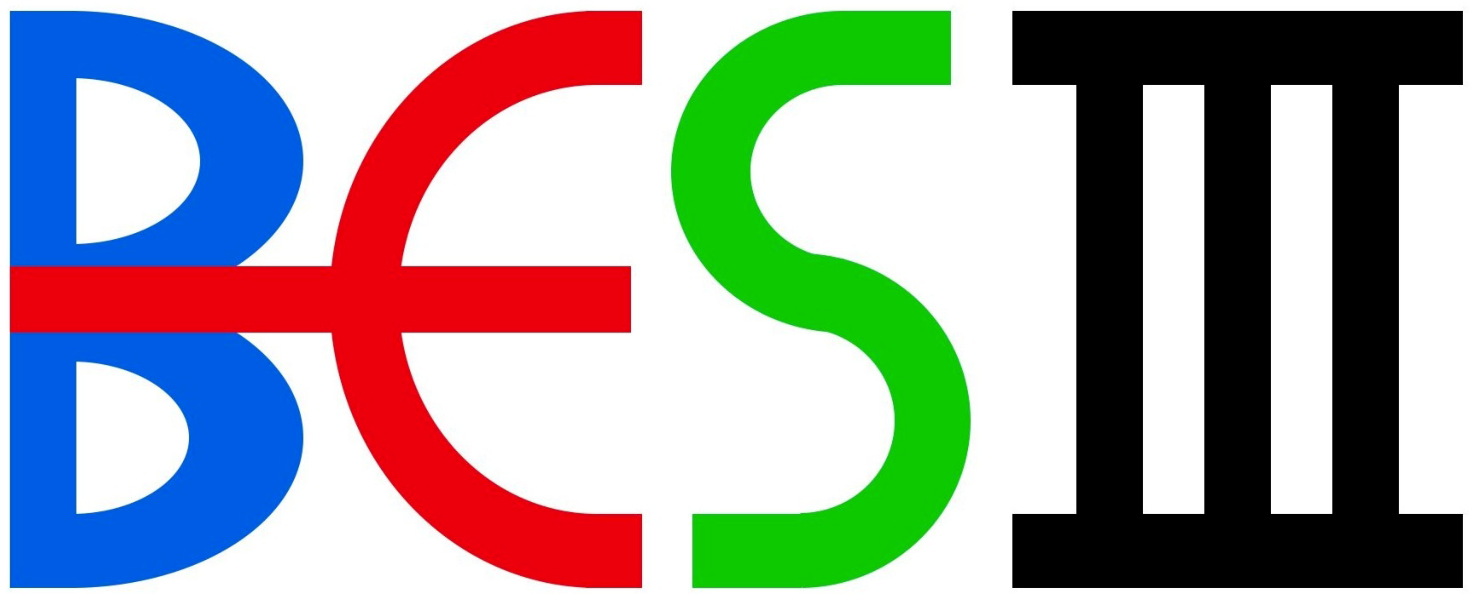}}

\collaboration{The BESIII collaboration}
\emailAdd{BESIII-publications@ihep.ac.cn}

\abstract{
By analyzing $e^{+}e^{-}$ collision data with an integrated
  luminosity of 7.9~fb$^{-1}$ collected with the BESIII detector at the
  center-of-mass energy of 3.773~GeV, the branching fraction of $D^+\to\tau^+\nu_{\tau}$ is determined as 
  $\mathcal{B}=(9.9\pm 1.1_\mathrm{stat}\pm 0.5_\mathrm{syst})\times10^{-4}$.
Using the most precise result $\mathcal{B}(D^+\to\mu^+\nu_{\mu})=(3.981\pm 0.079_\mathrm{stat}\pm0.040_\mathrm{syst})\times10^{-4}$~\cite{BESIII:2024kvt}, we determine $R_{\tau/\mu} = \Gamma(D^+\to\tau^+\nu_{\tau})/\Gamma(D^+\to\mu^+\nu_{\mu})= 2.49\pm0.31 $, achieving a factor of two improvement in precision compared to the previous BESIII result. This measurement is in agreement with the standard model prediction of lepton flavor universality within one standard deviation.

}



\maketitle
\flushbottom

\section{INTRODUCTION}
\label{sec:introduction}
\hspace{1.5em} 

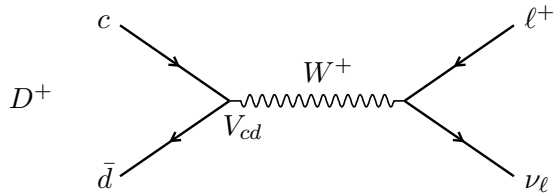
\begin{figure}[htbp]
  \centering
  \includegraphics[height=3cm]{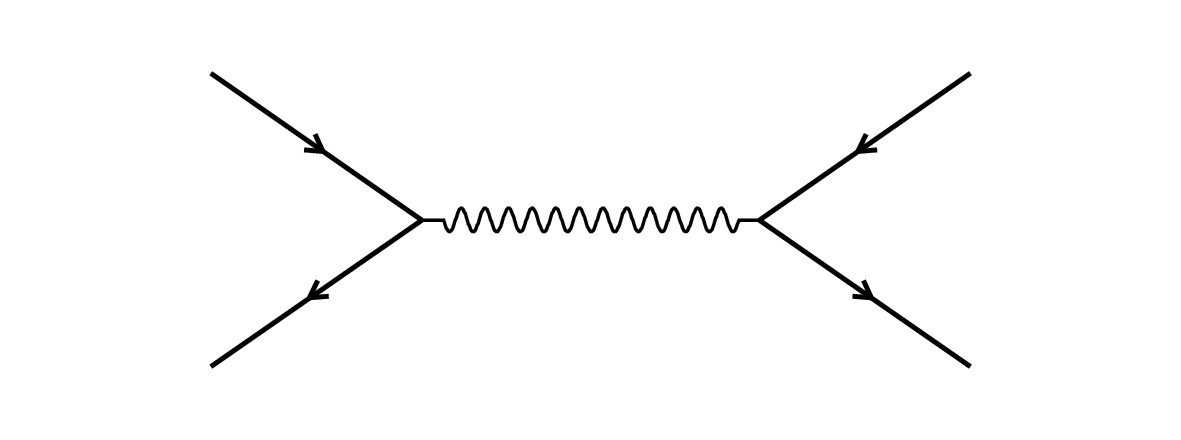}
  \put(-197,70){\bf $c$}
  \put(-197,10){\bf $\bar d$}
  \put(-230,40){\bf $D^+$}
  \put(-150,30){\bf $V_{cd}$}
  \put(-120,50){\bf $W^+$}
  \put(-37,70){\bf $\ell^+$}
  \put(-37,10){\bf $\nu_\ell$}
 \caption{Tree level Feynman diagram of the leptonic decay  of the $D^+$ meson. The
    CKM matrix element, $V_{cd}$, appears at the decay vertex.}
  \label{fig:diagram}
\end{figure}

Lepton Flavor Universality~(LFU), a fundamental assumption of the
  Standard Model~(SM), asserts that leptons from different generations
  possess identical coupling to gauge bosons~\cite{Ke:2023qzc}.
  The leptonic decays of the $D^+$ meson
  present an excellent opportunity for testing $\mu-\tau$ LFU.
  Figure~\ref{fig:diagram} shows the Feynman diagram of the
  $D^+ \to \ell^+\nu_\ell$~($\ell=e,\mu,\tau$) decays at tree level, in which the $c$ quark and the $\bar{d}$
  quark annihilate, followed by a weak current connecting to the system of the
  lepton $\ell^+$ and the corresponding flavored neutrino
  $\nu_\ell$. Throughout this paper, charge conjugate channels are implied.

Due to the isolation of the hadronic and leptonic systems,
the partial decay width of $D^+ \to \ell^+\nu_\ell$ to the lowest
order within the SM is
proportional to the product of the decay constant
$f_{D^+}$ and the Cabibbo-Kobayashi-Maskawa~(CKM) matrix element $|V_{cd}|$,
given in a simple form~\cite{Ke:2023qzc, PDG, PhysRevD.38.214}:
\begin{equation}
\small
\Gamma(D^+ \to \ell^+\nu_\ell)= \frac{G^2_F f^2_{D^+}} {8\pi}|V_{cd}|^2M^2_{\ell^{+}} M_{D^+}
    \left (1- \frac{M^2_{\ell^{+}}}{M^2_{D^+}}\right )^2,
\label{eq01}
\end{equation}
where $M_\ell$ and $M_{D^+}$ are the masses of the lepton $\ell$
and the $D^+$ meson, respectively.
The partial decay width is the branching fraction divided by the $D^+$ lifetime.
$G_{F}$ is the Fermi coupling constant, which is known to high precision.
From Eq.~\ref{eq01} and assuming LFU, the ratio of the branching fractions of $D^+\to\tau^+\nu_{\tau}$ and $D^+\to\mu^+\nu_{\mu}$
can be predicted with negligible uncertainty as 
\begin{equation}
\small
R_{\tau/\mu}= \frac{\Gamma(D^+\to\tau^+\nu_{\tau})}{\Gamma(D^+\to\mu^+\nu_{\mu})} =  \frac{M_{\tau}^2\left (1-\frac{M^2_\tau}{M^2_{D^+}}\right )^2} {M_{\mu}^2\left (1- \frac{M^2_\mu}{M^2_{D^+}}\right )^2} = 2.67.
\label{eq02}
\end{equation} 
It is important to note that this expression is independent of $G_F$, $|V_{cd}|$ and $f_{D^+}$, while the uncertainties associated with the higher-order corrections and $M_{\ell/D^+}$ are negligible.

Some new physics models, such as the
two-Higgs-doublet model~\cite{Fajfer:2015ixa,Akeroyd:2007eh}, mediated via
charged Higgs bosons, or the seesaw mechanism, which involves lepton mixing with
Majorana neutrinos~\cite{Branco:2011zb}, may lead to LFU violation.  LFU has been comprehensively tested in the $B$ sector by
the BaBar, LHCb, and
Belle Collaborations~\cite{BaBar:2012obs,BaBar:2013mob,LHCb:2015gmp,LHCb:2014vgu,Belle:2016fev, LHCb:2023zxo}.
However, tests of $\tau$-$\mu$ LFU in the charm sector are still limited by the $\mathcal{B}(D^+\to\tau^+\nu_\tau)$ measurement accuracy.
Previously, BESIII reported the most precise measurement of the
$D^+\to\mu^+\nu_\mu$ branching fraction~\cite{BESIII:2024kvt} and the observation of
$D^+\to\tau^+\nu_\tau$~\cite{Ablikim:2019rpl} via
$\tau^{+}\to\pi^+\bar{\nu}_\tau$ based on 2.9~fb$^{-1}$~\cite{Ablikim:2013ntc}
of $e^+e^-$ collision data taken at $\sqrt{s}=3.773$~GeV, resulting in a ratio $R_{\tau/\mu} = 3.21\pm0.64\pm0.43 $. Improved measurements are essential to obtain better knowledge of this ratio and to enable more sensitive LFU tests. For $D^+_{s}$ leptonic decays, the ratio of $D^+_{s}\to\tau^+\nu_\tau$ and $D^+_{s}\to\mu^+\nu_\mu$ has been reported as $R_{\tau/\mu} = 9.72\pm0.37$~\cite{BESIII:2021bdp}, which is in agreement with the SM prediction 9.75. Our analysis is complementary with the $D_s^+$  results in testing the effects of new physics, due to the symmetry of SU(3). Furthermore, the global fit that includes both $D^+$ and $D_s^+$ decays  may  impose more stringent constraints on new physics. 

In this paper, we report an improved measurement of
$\mathcal{B}(D^+\to\tau^+\nu_{\tau})$ and $R_{\tau/\mu}$,  with an
integrated luminosity of 7.9~fb$^{-1}$ data collected at $\sqrt{s}=3.773$~GeV  with the BESIII detector
in 2010, 2021 and 2022. The precision is improved by a factor
of two compared with the previous BESIII result~\cite{Ablikim:2019rpl}.
Furthermore, based on the measured partial decay width
of $D ^+ \to \ell^+\nu_\ell$\footnote{The partial decay width is calculated as the measured branching fraction divided
by the $D^+$ lifetime obtained through a global fit~\cite{PDG}.},
we determine $f_{D^+}$ by taking the $|V_{cd}|$ from the SM global
fit~\cite{PDG}, and also extract $|V_{cd}|$ taking as input 
$f_{D^+}$ from lattice quantum
chromodynamics~(LQCD) calculations~\cite{FlavourLatticeAveragingGroupFLAG:2021npn}.

\section{BESIII DETECTOR AND MONTE CARLO SIMULATION}
\label{sec:detector}
\hspace{1.5em}

The BESIII detector~\cite{Ablikim:2009aa} records symmetric $e^+e^-$ collisions 
provided by the BEPCII storage ring~\cite{Yu:IPAC2016-TUYA01}
in the center-of-mass energy range from 1.85 to 4.95~GeV,
with a peak luminosity of $1.1 \times 10^{33}\;\text{cm}^{-2}\text{s}^{-1}$ 
achieved at $\sqrt{s} = 3.773\;\text{GeV}$. 
BESIII has collected large data samples in this energy region~\cite{Ablikim:2019hff, EcmsMea, EventFilter}. The cylindrical core of the BESIII detector covers 93\% of the full solid angle and consists of a helium-based
 multilayer drift chamber~(MDC), a plastic scintillator time-of-flight
system~(TOF), and a CsI(Tl) electromagnetic calorimeter~(EMC), 
which are all enclosed in a superconducting solenoidal magnet
providing a 1.0~T magnetic field.
The solenoid is supported by an
octagonal flux-return yoke with resistive plate counter muon
identification modules interleaved with steel. 
The charged-particle momentum resolution at $1~{\rm GeV}/c$ is
$0.5\%$, and the 
${\rm d}E/{\rm d}x$
resolution is $6\%$ for electrons
from Bhabha scattering. The EMC measures photon energies with a
resolution of $2.5\%$ ($5\%$) at $1$~GeV in the barrel (end-cap)
region. The time resolution in the TOF barrel region is 68~ps, while
that in the end-cap region was 110~ps. The end-cap TOF
system was upgraded in 2015 using multigap resistive plate chamber
technology, providing a time resolution of
60~ps, which benefits 63\% of the data used in this analysis~\cite{etof}.

Monte Carlo (MC) simulated data samples produced with a {\sc
geant4}-based~\cite{geant4} software package, which
includes the geometric description of the BESIII detector and the
detector response, are used to determine detection efficiencies
and to estimate backgrounds. The simulation models the beam-energy spread and initial-state radiation (ISR) in the $e^+e^-$
annihilations with the generator {\sc
kkmc}~\cite{ref:kkmc}.
The inclusive MC sample includes the production of $D\bar{D}$
pairs (including quantum coherence for the neutral $D$ channels),
the non-$D\bar{D}$ decays of the $\psi(3770)$, the ISR
production of the $J/\psi$ and $\psi(3686)$ states, and the
continuum processes incorporated in {\sc kkmc}~\cite{ref:kkmc}.
All particle decays are modelled with {\sc
evtgen}~\cite{ref:evtgen} using branching fractions 
either taken from the
Particle Data Group~\cite{PDG}, when available,
or otherwise estimated with {\sc lundcharm}~\cite{ref:lundcharm}.
Final-state radiation
from charged final-state particles is incorporated using the {\sc
photos} package~\cite{photos2}. The leptonic decay $D^{+} \to \tau^+\nu_{\tau}$ is simulated with the SLN model~\cite{ref:evtgen}.

\section{EVENT SELECTION AND DATA ANALYSIS}
\label{sec:analysis}
\hspace{1.5em} 

The analysis uses the double-tag~(DT) technique~\cite{MARK-III:1987jsm}.
All the selection criteria follow those adopted in the previous BESIII study~\cite{Ablikim:2019rpl}. 
On the single-tag~(ST) side, the $D^-$ mesons are reconstructed via
the six hadronic decay modes $K^+\pi^-\pi^-$, $K^+\pi^-\pi^-\pi^0$, $K_S^0\pi^-$, $K_S^0\pi^-\pi^0$, and $K_S^0\pi^-\pi^-\pi^+$. 
On the signal side, the $\tau^+\to\pi^+\bar{\nu}_{\tau}$ process is used
to reconstruct $D^{+} \to \tau^+\nu_{\tau}$ and thereby measure $R_{\tau/\mu}$.

\begin{figure}[htbp]
  \centering
   \includegraphics[width=14 cm]{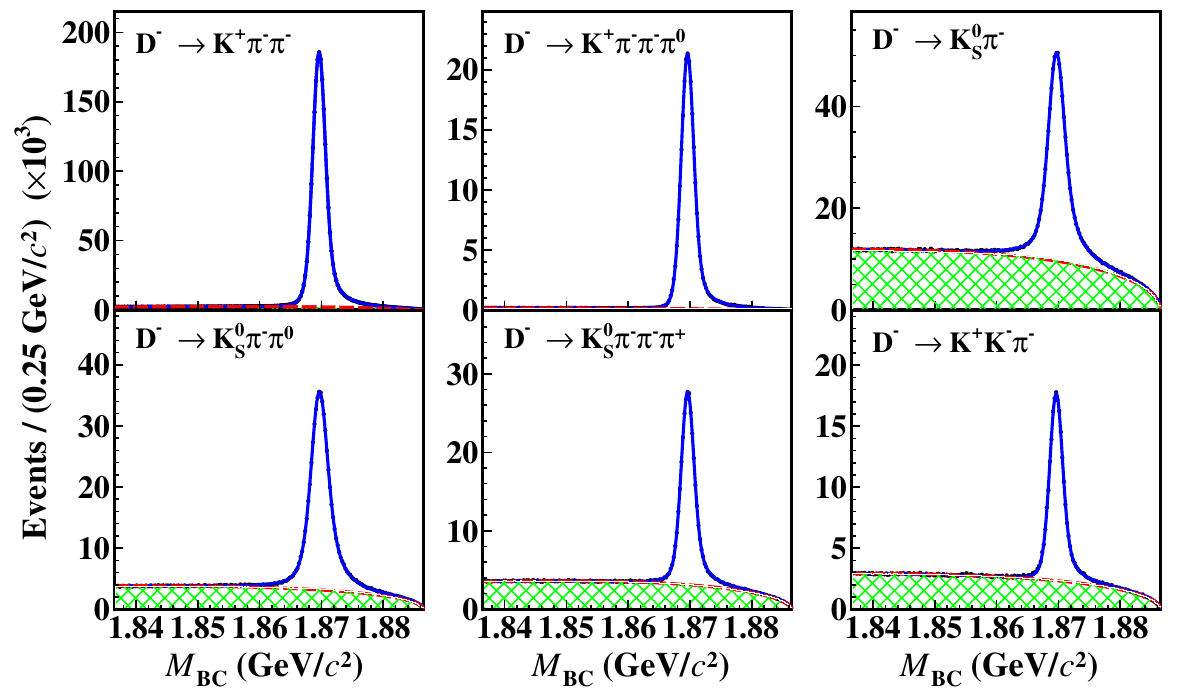}
  \caption{
Fits to the $M_{\rm BC}$ distributions of the ST $D^-$ candidates. The dots with error bars are data. The blue solid curves are the fit results. The red dashed curves are the fitted combinatorial backgrounds. The green hatched histograms are background events from the inclusive MC sample.}
  \label{fig:fit_MBC}
\end{figure}

On the ST side, to distinguish the $D^-$ mesons
from the combinatorial backgrounds, two variables, the beam-constrained mass $M_{\rm BC}\equiv \sqrt{E_{\rm beam}^2-|\vec{p}_{D^-}|^2}$ and energy difference $\Delta{E}\equiv E_{D^-}-E_{\rm beam}$ are defined, 
where $\vec{p}_{D^-}$ and $E_{D^-}$ are the total reconstructed momentum and energy of
the $D^-$ candidate in the $e^+e^-$ center-of-mass frame, and $E_{\rm beam}$ is the calibrated beam energy. For each ST decay mode, if more than one candidate satisfies the above requirements, the one with the smallest value of  $|\Delta{E}|$ is kept for further analyses. The ST yields for data are obtained by a fit to the $M_{\rm BC}$ distributions with MC based signal shapes convolved with double Gaussian functions.
The background shape is parametrized by an ARGUS function with free parameters, except for
the endpoint, which is fixed at 1.8865~GeV corresponding to $E_{\rm beam}$.
The fits are shown in Fig.~\ref{fig:fit_MBC}.
The ST efficiencies are determined by analyzing inclusive MC samples.
Those candidates with $M_{\rm BC}$ lying within (1.863, 1.877)~GeV/c$^2$ are retained for the subsequent analysis.
The $|\Delta E|$ requirements, ST yields and the ST efficiencies are shown in Table~\ref{tab:ST}.

\begin{table}[hbtp]
  \begin{center}
    \caption{The $\Delta E$ requirements, ST yield
      $(N_{\rm tag}^{\rm ST})$ and ST efficiencies $(\epsilon_{\rm tag}^{\rm ST})$.
      }
    \begin{tabular}{lccccc}
      \hline
      \hline
      Tag mode  & $\Delta E$ (MeV)  &$N_{\rm tag}^{\rm ST}$ & {$\epsilon_{\rm tag}^{\rm ST}(\%)$} \\ 
      \hline
      $K^+\pi^-\pi^-$              &(-25,24) & 2164074$\pm$1571 & 51.17$\pm$0.01 \\ 
      $K^+\pi^-\pi^-\pi^0$      &(-57,46) & 689042$\pm$1172  & 25.50$\pm$0.01\\ 
      $K_S^0\pi^-$                &(-25,26) & 250437$\pm$524  & 50.63$\pm$0.02\\ 
      $K_S^0\pi^-\pi^0$         &(-62,49) & 558495$\pm$930   & 26.28$\pm$0.01 \\ 
      $K_S^0\pi^-\pi^-\pi^+$  &(-28,27) & 300519$\pm$669   & 28.97$\pm$0.01 \\ 
      $K^+K^-\pi^-$               &(-24,23) & 187379$\pm$541   & 41.06$\pm$0.02 \\ 
      \hline
      \hline
    \end{tabular}
    \label{tab:ST}
  \end{center}
\end{table}

After the ST event is identified, the remaining tracks which have not been used in the ST-side reconstruction are used to select candidates for the signal process $D^{+} \to \tau^+\nu_{\tau}$,$\tau^+\to\pi^+\bar{\nu}_{\tau}$. Since the neutrinos are not detected, only one charged pion needs to be selected in a candidate event. Hence, we require that there must be only one additional track with opposite charge to the ST side. The selection criteria of the pions are identical to those of the ST side and have to match a shower in the EMC additionally. Since pions and muons are both charged particles with similar masses, the pion selection 
 also accepts muon tracks with comparable efficiency ($>$90\%). However, the deposited energy in the EMC ($E_{\rm EMC}$) of most muons is less than $300$~MeV, which is used to separate muon from pion. To increase the signal  significance, we partition the data into a $\pi$-like sample  with $E_{\rm EMC}>300 $~MeV  and a  $\mu$-like sample  with $E_{\rm EMC}\le300$~MeV.

Four additional requirements are applied to further suppress background processes.
These requirements are set from studies of  MC data and the systematic uncertainties are assessed from studies of control samples in the data.
(1)~Processes with misidentified electrons backgrounds can contaminate our signal sample, for example the semileptonic decay $D^+\to K_L^0 e^+ \nu_{e}$ when the $K_L^0$ is missed. The quantity $E_{\rm EMC}/p$ is distributed around unity for background processes in this category, where $p$ is the momentum of the track reconstructed in the MDC.   Therefore, we require $E_{\rm EMC}/p<0.95$ for events in the $\pi$-like sample.
(2)~In order to suppress backgrounds with additional neutral particles, such as  $D^+\to\eta\pi^+$, we demand $E_{\gamma,\text{max}}<300$~MeV for both samples, where $E_{\gamma,\text{max}}$ is
the maximum energy of all EMC showers not used in the reconstruction of either the ST or DT side.
(3)~We require $|\!\cos\theta_{\rm miss}|<0.75(0.97)$ for the $\pi$-like ($\mu$-like) sample to ensure that
the missing track points to the active region of the detector, where $\cos\theta_{\rm miss}$ is the polar angle of the missing momentum.
(4)~In order to suppress background from  $D^+\to K_L^0\pi^+$ decays we impose $\alpha > 45^\circ(25^\circ)$ for the $\pi$-like ($\mu$-like) sample, where $\alpha$ is the opening angle between the missing track and
the most energetic unassigned shower.

\section{DETERMINATION OF THE BRANCHING FRACTION}
\hspace{1.5em}

\begin{figure}[htbp]
  \centering
   \includegraphics[width=7cm]{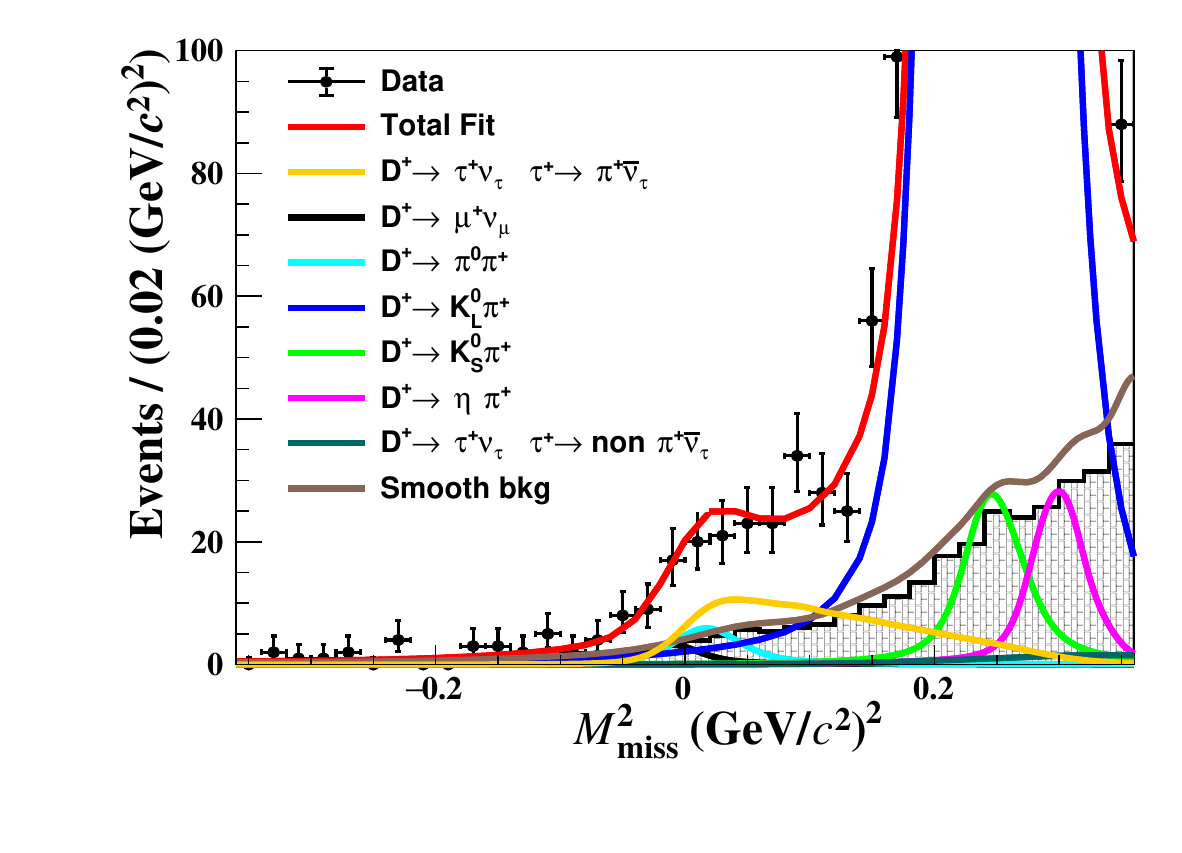}
    \put(-150,45){(a)}
  \includegraphics[width=7cm]{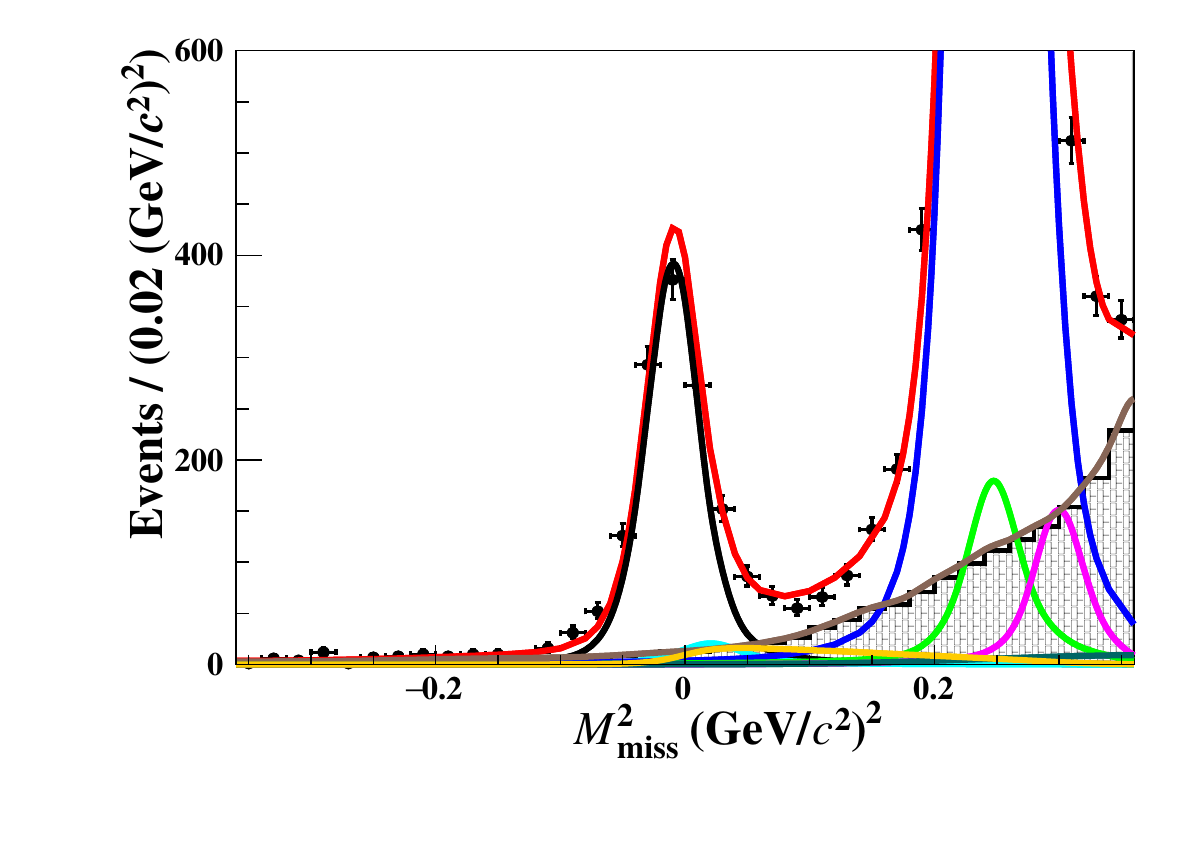}
   \put(-150,45){(b)}
  \caption{
Fits to the distributions of $M^2_{\rm miss}$ for $\pi$-like (a) and $\mu$-like (b)
cases. 
All of the rest of the colored lines correspond to fits to background components.
}
  \label{fig:fit_MM2}
\end{figure}

The quantity  missing-mass squared~(${M_{\rm miss}^2}$) is defined to provide sensitivity to the 
missing neutrino: 
\begin{equation}
{M_{\rm miss}^2}=(E_{\rm cm}-E_{D^-}-E_{\rm \pi})^2-(\vec{p}_{\rm cm}-\vec{p}_{D^-}-\vec{p}_{\rm \pi})^2,
\label{def:MM2}
\end{equation}
where $E_{\rm cm}$~($\vec{p}_{\rm cm}$) is the center-of-mass energy~(momentum),
$E_{\rm \pi}$~($\vec{p}_{\rm \pi}$) is the energy~(momentum) of the pion,
and $E_{D^-}$~($\vec{p}_{D^-}$) is the energy~(momentum) of the $D^-$ tag.
An unbinned simultaneous maximum likelihood fit is performed on the $M_{\rm miss}^2$ distributions for both the $\pi$-like and $\mu$-like samples,
shown in Fig.~\ref{fig:fit_MM2}, to obtain the yields. The branching fraction of signal process is a common parameter in the fit and is obtained through 
\begin{equation}
  \mathcal{B}(D^+\to\tau^+\nu_{\tau}) = \frac{N_{\tau\nu}}{\sum_{i}{N^{\rm ST}_{i}}\epsilon^{\rm DT}_{i,\tau\nu}/\epsilon^{\rm ST}_{i}},
  \label{abs:bf}
\end{equation}  
where $N_{\tau\nu}$ is the DT yields for all tag modes, $\epsilon^{\rm DT}_{i,\tau\nu}$ is the signal selection efficiencies without $B(\tau^+\to\pi^+\bar{\nu}_{\tau})$ shown in Table~\ref{tab:eff_DT_signal}, $N^{\rm ST}_{i}$ and $\epsilon^{\rm ST}_{i}$ are the numbers of the ST events and ST efficiencies for the ST mode $i$.  In extracting the DT yields for each tag mode the $D^+\to\tau^+\nu_\tau$ ($\tau^+\to\pi^+\bar{\nu}_\tau$) signal shape is taken from the MC sample.
The peaking-background shapes of $D^+\to\pi^0\pi^+$ and $D^+\to K_L^0\pi^+$ are extracted from data-based control samples.  All the other shapes are determined using inclusive MC simulation and constructed using kernel estimation PDF~\cite{Cranmer:2000du} in RooFit~\cite{Canal:2015bjt}.  Additionally, the yields of the signal process $D^+\to\tau^+\nu_\tau$ ($\tau^+\to\pi^+\bar{\nu}_\tau$) and the significant peaking background $D^+\to K_L^0\pi^+$ are floated in the fit. The other smaller peaking backgrounds, $D^+\to\pi^0\pi^+$, $D^+\to\eta\pi^+$, and $D^+\to K_S^0\pi^+$ are fixed based on their known branching fractions quoted from the PDG~\cite{PDG} and $D^+\to\mu^+\nu_\mu$ quoted from BESIII measurement~\cite{BESIII:2024kvt}. The efficiencies obtained by the MC samples. The normalization of the remaining relatively smooth backgrounds is floated, but the relative ratio between $\pi$-like and $\mu$-like samples is constrained, based on study of the inclusive MC sample.  The $D^+\to K_S^0\pi^+$ control sample is used to estimate the difference of data and MC to correct the ratio.

\begin{table}[hbtp]
  \begin{center}
    \caption{The signal efficiencies $(\epsilon_{\pi(\mu)} = \epsilon_{\pi(\mu)}^{\rm DT} / \epsilon_{\rm tag}^{\rm ST})$
      for the $\pi(\mu)$-like selection. The branching fractions of the sub-particle ($\tau^+\to\pi^+\nu_\tau$) decays
      are not included. The uncertainties are statistical only.}
    \begin{footnotesize}
        \begin{tabular}{lccc}
      \hline
      \hline
      Tag mode  & $\epsilon_{\pi}$ (\%)& $\epsilon_{\mu}$ (\%)\\ 
      \hline
       \hline
      $D^-\to K^+\pi^-\pi^-$               & 39.54$\pm$0.23 &25.00$\pm$0.19\\ 
      $D^-\to K_S^0\pi^-$                 & 40.48$\pm$0.24 &25.60$\pm$0.19\\ 
      $D^-\to K^+\pi^-\pi^-\pi^0$        & 38.94$\pm$0.33 &25.45$\pm$0.25\\ 
      $D^-\to K_S^0\pi^-\pi^0$           & 40.19$\pm$0.32 &25.07$\pm$0.26\\ 
      $D^-\to K_S^0\pi^-\pi^-\pi^+$    & 37.52$\pm$0.31 &24.19$\pm$0.25\\ 
      $D^-\to K^+K^-\pi^-$                 & 37.74$\pm$0.26 &23.14$\pm$0.20\\ 
      \hline
      \hline
    \end{tabular}
     \end{footnotesize}
    \label{tab:eff_DT_signal}
  \end{center}
\end{table}

The signal yield is $283\pm32$ with a statistical significance of $6.5\sigma$. 
The branching fraction of $D^+\to\tau^+\nu_{\tau}$ is determined to
be $\mathcal{B}(D^+\to\tau^+\nu_{\tau})=(9.9\pm 1.1)\times 10^{-4}$,
where the uncertainty is statistical only.
As a cross check, we repeat the fit with the yields of $D^+\to\mu^+\nu_{\mu}$
floated, and obtain
$\mathcal{B}(D^+\to\tau^+\nu_{\tau})=(1.01\pm 0.13)\times 10^{-3}$
and $\mathcal{B}(D^+\to\mu^+\nu_{\mu})=(3.89\pm 0.12)\times 10^{-4}$.
The latter is consistent with the PDG value of $(3.74\pm 0.17)\times 10^{-4}$~\cite{PDG}.

\section{SYSTEMATIC UNCERTAINTIES}
\hspace{1.5em}
Benefiting from the DT technique, we assume that all systematic
uncertainties due to the reconstruction of particles on the ST side 
largely cancel. The systematic uncertainties are dominated by the selection criteria on the DT side, the simultaneous fit, and the knowledge of $\mathcal{B}(D^+\to\mu^+\nu_{\mu})$.  These are summarized in Table~\ref{tab:absBF_sys}.
\begin{table}[htbp]
\begin{center}
  \caption{Systematic uncertainties for the branching fraction measurement.}
  \begin{footnotesize}
  \centering
  \begin{tabular}{cc}
    \hline
    \hline
    Source   &Sys. Uncertainty(\%) \\
    \hline
    $E_{\rm EMC}$ requirement & 1.1\\
    Smooth background shape&1.8\\
    $\pi^+$ tracking & 1.0\\
    $\pi^+$ PID & 1.0\\
    $N_{\rm ST}$ & 0.1\\
    Requirement on $E_{\rm EMC}/p$ & 1.1\\
    Requirement on $E_{\gamma,\text{max}}$ & 1.2\\
    Requirements on $|\!\cos\theta_{\rm miss}|$ and $\alpha$ & 2.9\\
    Tag bias & 0.1\\
    Normalizations of small peaking BKG & 1.5\\
    Relative size ratio of smooth BKG & 1.5\\
    Signal shape of $D^+\to\tau^+\nu_{\tau}$ & 0.5\\
    $\mathcal B(D^+\to\mu^+\nu_{\mu})$ & 2.1\\
    $\mathcal B(\tau^+\to\pi^+\bar{\nu}_\tau)$ & 0.5\\
    \hline
    Total    	 	        & 5.2\\
    \hline
    \hline
  \end{tabular}
  \end{footnotesize}
  \label{tab:absBF_sys}
  \end{center}
\end{table}

The systematic uncertainties associated with the  $E_{\rm EMC}$, $E_{\rm EMC}/p$ and
$E_{\gamma,\text{max}}$ requirements are studied using a $D^+\to K_S^0\pi^+$ control
sample.  Those associated  with the $|\!\cos\theta_{\rm miss}|$ and missing cone $\alpha$ requirements are estimated by a control sample of  $D^0\to K^-e^+\nu_{e}$ decays. The difference in efficiency  of each requirement between data and MC simulation is assigned to be the corresponding systematic uncertainty.
To assess the uncertainty associated with the simultaneous fit, we change the MC components for example exclude the non-$D\bar{D}$ process to estimate the effect on the smooth background shapes. We also vary the $E_{\rm EMC}$ requirement by  $\pm$50~MeV to account for the difference in resolution between  data and MC.  We vary the normalizations of the small peaking backgrounds $D^+\to K_S^0\pi^+$, $D^+\to\pi^0\pi^+$ and $D^+\to\eta\pi^+$ by $\pm1\sigma$ of their known branching fraction. The ratio of the smooth background contribution between $\pi$-like and $\mu$-like samples in the $M_{\rm miss}^2$fit has been corrected by $D^+\to K_S^0\pi^+$ control sample and the error is assumed as the systematic uncertainty.  
We vary the fixed branching fraction of $\mathcal{B}(D^+\to\mu^+\nu_{\mu})$~\cite{BESIII:2024kvt} by
$\pm1\sigma$ to estimate the corresponding systematic uncertainty.
\section{SUMMARY}
\label{sec:summary}
\hspace{1.5em}
Based on $e^+e^-$ collision data corresponding to an integrated luminosity of
7.9~fb$^{-1}$ collected with the BESIII detector at $\sqrt{s} = 3.773$~GeV,
we measure $\mathcal{B}(D^+\to\tau^+\nu_{\tau})=(9.9\pm1.1_\mathrm{stat}\pm0.5_\mathrm{syst})\times10^{-4}$,
while fixing $\mathcal{B}(D^+\to\mu^+\nu_{\mu})=(3.981\pm0.079_\mathrm{stat}\pm0.040_\mathrm{syst})\times10^{-4}$~\cite{BESIII:2024kvt}. This result is consistent with the previous BESIII measurement of $(1.20\pm0.24_\mathrm{stat}\pm0.12_\mathrm{syst})\times10^{-3}$~\cite{Ablikim:2019rpl} and is twice as precise. The ratio $R_{\tau/\mu}$ is
determined to be $2.49\pm0.31$, which is consistent with the SM prediction of 2.67 within one standard deviation.
\begin{table}[hbtp]
  \begin{center}
   \begin{footnotesize}
    \caption{External input parameters with uncertainties.}
    \begin{tabular}{lc}
      \hline
      \hline
      Parameter & Value\\ 
      \hline
       \hline
$m_\mu$ &$0.10566(24)$ GeV\\ 
$m_\tau$& $1.77686(12)$ GeV\\
$M_{D^+}$& $1.86966(5)$ GeV\\
$G_F$& $1.1663787(6)\times10^{-5}$ GeV$^{-2}$\\
$\tau_{D^+}$& $1.033(5)$ ps \\
      \hline
      \hline
    \end{tabular}
    \label{tab:parameter}
     \end{footnotesize}
  \end{center}
\end{table}
\newpage
Combing the measured branching fraction with the world average value of $G_F$, $m_\ell$, $M_{D^+}$ and the $D^+$ lifetime ($\tau_{D^+}$) shown
in Table~\ref{tab:parameter}, we obtain $f_{D^+}|V_{cd}|=(45.9\pm 2.5_\mathrm{stat}\pm 1.2_\mathrm{syst}\pm 0.1_\mathrm{input})$~MeV. Here the third uncertainty arises from the  knowledge of $\tau_{D^+}$~(0.2\%). Taking the CKM matrix element $|V_{cd}| = 0.22486\pm0.00067$ from the global SM fit~\cite{PDG} we obtain 
 $f_{D^+}=(204\pm11_\mathrm{stat}\pm5_\mathrm{syst}\pm 1_\mathrm{input})$~MeV, which is in agreement with the value of $f_{D^+} = (212.1\pm0.7)$~MeV from recent LQCD calculations~\cite{FlavourLatticeAveragingGroupFLAG:2021npn}.  Alternatively, taking the LQCD value of the decay constant as input, we obtain $|V_{cd}|=(0.216\pm0.012_\mathrm{stat}\pm0.006_\mathrm{syst}\pm 0.001_\mathrm{input})$, which is in agreement with the value from the global SM fit.  \\

\hspace{1.5em}
\textbf{Acknowledgement}

The BESIII Collaboration thanks the staff of BEPCII and the IHEP computing center for their strong support. This work is supported in part by National Key R\&D Program of China under Contracts Nos. 2023YFA1606000, 2020YFA0406400, 2020YFA0406300; National Natural Science Foundation of China (NSFC) under Contracts Nos. 11635010, 11735014, 11935015, 11935016, 11935018, 12025502, 12035009, 12035013, 12061131003, 12192260, 12192261, 12192262, 12192263, 12192264, 12192265, 12221005, 12225509, 12235017, 12361141819, 12305105; the Chinese Academy of Sciences (CAS) Large-Scale Scientific Facility Program; the CAS Center for Excellence in Particle Physics (CCEPP); Joint Large-Scale Scientific Facility Funds of the NSFC and CAS under Contract No. U2032104, U1832207; The Excellent Youth Foundation of Henan Scientific Committee under Contract No.~242300421044; Natural Science Foundation of Shandong Province under Grants No.~ZR2023QA119; 100 Talents Program of CAS; The Institute of Nuclear and Particle Physics (INPAC) and Shanghai Key Laboratory for Particle Physics and Cosmology; German Research Foundation DFG under Contracts Nos. FOR5327, GRK 2149; Istituto Nazionale di Fisica Nucleare, Italy; Knut and Alice Wallenberg Foundation under Contracts Nos. 2021.0174, 2021.0299; Ministry of Development of Turkey under Contract No. DPT2006K-120470; National Research Foundation of Korea under Contract No. NRF-2022R1A2C1092335; National Science and Technology fund of Mongolia; National Science Research and Innovation Fund (NSRF) via the Program Management Unit for Human Resources \& Institutional Development, Research and Innovation of Thailand under Contracts Nos. B16F640076, B50G670107; Polish National Science Centre under Contract No. 2019/35/O/ST2/02907; Swedish Research Council under Contract No. 2019.04595; The Swedish Foundation for International Cooperation in Research and Higher Education under Contract No. CH2018-7756; U. S. Department of Energy under Contract No. DE-FG02-05ER41374.

%

\newpage
M.~Ablikim$^{1}$, M.~N.~Achasov$^{4,c}$, P.~Adlarson$^{76}$, O.~Afedulidis$^{3}$, X.~C.~Ai$^{81}$, R.~Aliberti$^{35}$, A.~Amoroso$^{75A,75C}$, Y.~Bai$^{57}$, O.~Bakina$^{36}$, I.~Balossino$^{29A}$
% ...
\\
\vspace{0.2cm}
(BESIII Collaboration)\\
\vspace{0.2cm} {\it
$^{1}$ Institute of High Energy Physics, Beijing 100049, People's Republic of China\\
% ...
$^{3}$ Bochum  Ruhr-University, D-44780 Bochum, Germany\\
$^{4}$ Budker Institute of Nuclear Physics SB RAS (BINP), Novosibirsk 630090, Russia\\
% ...
$^{29}$ INFN Sezione di Ferrara, (A)INFN Sezione di Ferrara, I-44122, Ferrara, Italy; (B)University of Ferrara,  I-44122, Ferrara, Italy\\
% ...
$^{35}$ Johannes Gutenberg University of Mainz, Johann-Joachim-Becher-Weg 45, D-55099 Mainz, Germany\\
$^{36}$ Joint Institute for Nuclear Research, 141980 Dubna, Moscow region, Russia\\
% ...
$^{57}$ Southeast University, Nanjing 211100, People's Republic of China\\
% ...
$^{75}$ University of Turin and INFN, (A)University of Turin, I-10125, Turin, Italy; (B)University of Eastern Piedmont, I-15121, Alessandria, Italy; (C)INFN, I-10125, Turin, Italy\\
$^{76}$ Uppsala University, Box 516, SE-75120 Uppsala, Sweden\\
% ...
$^{81}$ Zhengzhou University, Zhengzhou 450001, People's Republic of China\\

\vspace{0.2cm}
% ...
$^{c}$ Also at the Novosibirsk State University, Novosibirsk, 630090, Russia\\
% ...

}

\end{document}